\documentclass[epj]{svjour}
\usepackage{latexsym}
\usepackage{graphicx} 
\usepackage{amsmath}
\usepackage{amssymb}
\begin{document}
\title{Classical Spin Liquid Properties of the Infinite-Component Spin Vector
Model on a Fully Frustrated Two Dimensional Lattice}
\author{Benjamin Canals\inst{1} \and D. A. Garanin\inst{2}% etc
}                     % Do not remove
\institute{
Laboratoire Louis N\'eel, CNRS, 25 avenue
des Martyrs, Boite Postale 166, 38042 Grenoble Cedex 9, France 
\and 
Max-Planck-Institut f\"ur Physik komplexer Systeme, 
N\"othnitzer Strasse 38, D-01187 Dresden, Germany}
\date{Received: date / Revised version: date}
\abstract{
Thermodynamic quantities and correlation functions (CFs) of the classical
antiferromagnet on the checkerboard lattice are studied for the exactly solvable
infinite-component spin-vector model, $D \to \infty$.
In contrast to conventional two-dimensional magnets with continuous symmetry
showing extended short-range order at distances smaller than the
correlation length, $r \lesssim \xi_c \propto \exp(T^*/T)$,  correlations in 
the checkerboard-lattice
model decay already at the scale of the lattice spacing due to
the strong degeneracy of the ground state characterized by a macroscopic
number of strongly fluctuating local degrees of freedom.  
At low temperatures, spin CFs decay as $\langle {\bf S}_0 {\bf S}_r \rangle
\propto 1/r^2$ in the range $a_0 \ll r \ll \xi_c \propto T^{-1/2}$, where $a_0$ is
the lattice spacing.
Analytical results for the principal thermodynamic quantities in our model are 
very similar with MC simulations, exact and analytical results for the classical 
Heisenberg model ($D=3$) on the pyrochlore lattice.
This shows that the ground state of the infinite-component spin vector model 
on the checkerboard lattice is a classical spin liquid. 
\PACS{
      {PACS-key}{discribing text of that key}   \and
      {PACS-key}{discribing text of that key}
     } % end of PACS codes
} %end of abstract
\maketitle
\section{introduction}
\label{secIntroduction}

Frustrated magnets are particularly interesting as they often involve
unusual low temperature magnetic behaviors.
Depending on the nature of interactions, the local connectivity, spin 
dimensionality, these systems stabilize unconventional magnetic
ground states such as non collinear N\'eel orders, topological spin glasses, 
classical or quantum spin liquids, spin ices \cite{diep94}.
These observations are both experimental and theoretical eventhough
some of them, like topological spin glasses, are still conjectures.
Among these ground states, two families have a striking property.
Spin liquids, as well as spin ices, possess a residual entropy at 
zero temperature.
This anomaly and the paradox it raises when considering the third
principle of thermodynamics, are major points that people have
tried to answer.

Theoretical studies of frustrated antiferromagnets started half
a century ago with the exact solution of the triangular Ising antiferromagnet
by Wannier \cite{wannier50}.
In the following years, many lattices have been identified where spin models, 
from Ising symmetry ($Z_2$)
to Heisenberg symmetry (O(3) or SU(2)), have disordered ground states.
Among them, two have attracted a lot of interest : the  
kagom\'e lattice and the pyrochlore lattice \cite{diep94}.
The first one is a two dimensional arrangement of corner sharing triangles,
while the latter is a three dimensional structure of corner sharing
tetrahedra.
When considering antiferromagnetic nearest neighbour interactions, they both
display spin liquid like behaviors, well characterized theoretically.
Spin-spin correlations functions are exponentially decaying at 
finite temperatures, even sometimes at $T=0$, in both classical and
quantum cases.

Recently, another lattice has received attention as it can be described
as the two dimensionnal analog of the pyrochlore lattice : the checkerboard
lattice.
One very interesting point in this lattice is that its geometry mimics
the local environment of each site of the pyrochlore lattice, but is
much more simpler as it is two dimensional, and based on a square lattice
structure.
This has allowed Lieb and Schupp \cite{lieb99} to establish exact results in the 
quantum case ($S=1/2$) showing that the
spectrum in this system is very peculiar as ground states are necesseraly
singlets in finite size clusters.
For $S \geq 1$, it has been shown numerically \cite{canalswave00} within an
$1/S$ expansion that the ground state may have a local magnetization at zero 
temperature while for $S=1/2$ it was proposed that the system orders in a
valence bond solid ground state.
A recent work \cite{moessner2001} derived a phase diagram of the ground states for varying $S$ 
stating that this system is most likely ordered for all $S$.
Exact diagonalization of finite clusters \cite{fouet2001} for $S=1/2$ reached the same
conclusion and define the ground state as a 4-spin valence bond solid.
A perturbative expansion also found a 4-spin valence bond solid \cite{elhajal2001} but
with a different unit cell.
There are now strong evidences that for small $S$ this system should have
long range order in a valence bond solid order parameter.

In this paper, we look for an exact solution on this lattice in the
opposite limit of high spin dimensionality.
Therefore, we study a classical Hamiltonian that corresponds to
the generalisation of the Heisenberg model with $D$-component spin
vectors \cite{sta68prl,sta74}
%
%\makebox{\it \bf dham}
%
\begin{equation}\label{dham}
{\cal H} = - \frac{1}{2} \sum_{rr'} J_{rr'} {\bf s}_r \cdot {\bf s}_{r'}  ,  \qquad |{\bf s}_r|=1
\end{equation}
and taking the limit $D\to\infty$.
In this limit the problem becomes exactly solvable for all lattice 
dimensionalities, $d$, and the partition function of the system 
coincides \cite{sta68pr} with that of 
the spherical model. \cite{berkac52,joy72pt}
The $D=\infty$ model properly accounts for the profound role played, especially in low dimensions, by the Goldstone or would be Goldstone modes.
At the same time,  the less significant effects of the critical fluctuation coupling leading, e.g., to the quantitatively different nonclassical critical indices, die out in the limit $D\to\infty$.
Thus this model is a relatively simple yet a powerful tool for classical spin systems.
It should not be mixed up with the $N$-flavour generalization of the quantum $S=1/2$ model 
\cite{auearo8888}  in the limit $N\to\infty$, including its $1/N$ expansion. \cite{sac92,timgirhen98}
The $N$-component nonlinear sigma model (see, e.g.,
Refs.~\cite{chahalnel8889}, as well as Ref.~\cite{chubsacye94,chubsta98} for the $1/N$ expansion) 
is a quantum extension of  Eq. (\ref{dham}) in the long-wavelength region at low temperatures.
Effective free energies for the $n$-component order parameter appear, instead of Eq. (\ref{dham}), in conventional theories of critical phenomena. 
Using them for the $1/n$ expansion (see, e.g., Ref.~\cite{ma73}) is a matter of taste.
While yielding the same results for the critical indices as the lattice-based $1/D$ expansion,
\cite{abe7273,abehik7377,okamas78} it misses the absolute values of the nonuniversal quantities.

In the following, we give the solution of this model
on the checkerboard lattice and show that its ground state is a
classical spin liquid.
Particularly, it is shown that many properties are similar 
to the pyrochlore lattice case \cite{canalsgaranin00}
despite the difference of lattice dimensionality.
This strongly suggests that these frustrated systems are mainly driven
by their local environment, i.e by their topological frustration, 
at least for $D\to\infty$.
The rest of this article is organized as follows.
In Sec. \ref{secStructure} the structure of the checkerboard lattice 
and its collective spin variables are described.
In Sec. \ref{secDinfinity} the formalism of the $D=\infty$ 
model is tailored for the checkerboard lattice. 
The diagrams of the classical spin diagram technique that 
do not disappear in the limit $D\to\infty$ are summed up.
The general analytical expressions for the thermodynamic functions 
and spin CFs for all temperatures are obtained.
In Sec. \ref{secThermod} the thermodynamic quantities of the checkerboard
antiferromagnet are calculated and
compared with MC simulation results as well as exact and analytical results previously
obtained on the pyrochlore lattice in the whole temperature range.
In Sec. \ref{secCFs} the real space correlation functions are computed.
We finally discuss our results in Sec. \ref{secDiscussion} and conclude.

\section{Lattice structure and the Hamiltonian}
\label{secStructure}

\begin{figure}
\includegraphics[width=7.5cm]{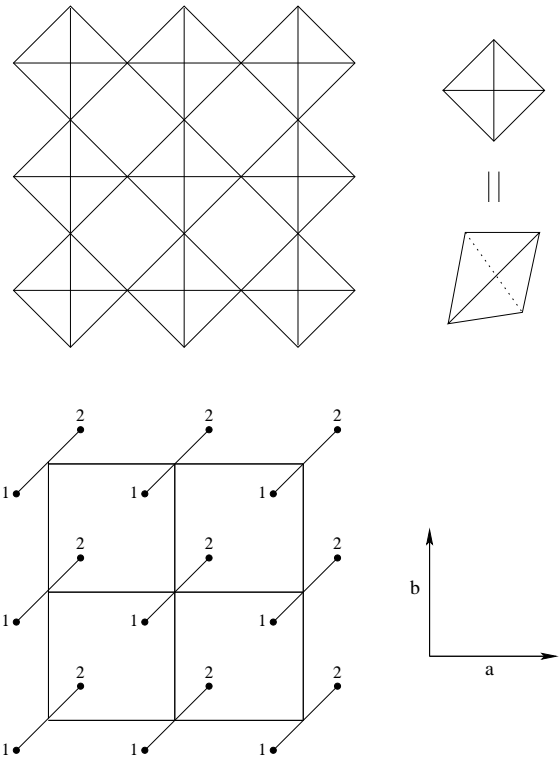}%
\caption{The checkerboard lattice. It can be pictured as a square lattice
of tetraheda (top). Locally, it reproduces the same environment
as the three dimensional pyrochlore lattice. In our
calculations, it has been described by a square 
lattice with a 2-spins unit cell (bottom).}
\label{fig-check-and-square}
\end{figure}

Checkerboard lattice shown in Fig.\ref{fig-check-and-square} consists 
of corner-sharing tetrahedra.
Each node of the corresponding Bravais lattice (i.e.,  each elementary
tetrahedron in Fig.\ref{fig-check-and-square}) is numbered by 
$i,j=1,\ldots, N$.
Each site of the elementary tetrahedron is labeled by the index $l=1,2$.
It is convenient to use the dimensionless units in which the interatomic
distance equals $1/2$ and hence the lattice period equals $1$.
The tetrahedra numbered by $i,j=1,\ldots, N$ can be obtained from each other by
the translations
%\makebox{\it \bf transl}
%
\begin{equation}
\label{transl}
{\bf r}_j^l =  {\bf r}_i^l + n_u {\bf u} + n_v {\bf v} ,
\end{equation}
where $ {\bf r}_i^l$ is the position of a site on the lattice, $n_u$ and $n_v$
are integers, ${\bf u}$  and ${\bf v}$ are the elementary translation vectors 
(lattice periods), and
%\makebox{\it \bf defuv}
%
\begin{equation}\label{defuv}
{\bf u}        = ( 1, 0 ), 
\qquad {\bf v} = ( 0, 1 ) .
\end{equation}

To facilitate the diagram summation in the next section, it is convenient to 
put the Hamiltonian (\ref{dham}) into a diagonal form.
First, one goes to the Fourier representation according to
%\makebox{\it \bf fourier}
%
\begin{equation}\label{fourier}
{\bf s}_{\bf q}^l = \sum_i {\bf s}_i^l e^{-i {\rm q\cdot r}_i^l}, 
\qquad {\bf s}_i^l = \frac 1N \sum_{\bf q} {\bf s}_{\bf q}^l e^{i {\rm q\cdot r}_i^l}, 
\end{equation}
where the wave vector ${\bf q}$ belongs to the square Brillouin zone
$\rm{Bz}_{kx,ky} = [- \pi , \pi] \times [- \pi , \pi]$
(see Fig. \ref{fig-check-and-square}).
The Fourier-transformed Hamiltonian reads
%\makebox{\it \bf dhamq}
%
\begin{equation}\label{dhamq}
{\cal H} = \frac{1}{2N} \sum_{ll'\bf q} V_{\bf q}^{ll'} {\bf s}_{\bf q}^l 
{ \cdot \rm s}_{-{\bf q}}^{l'} , 
\end{equation}
where the interaction matrix is given by
%\makebox{\it \bf Vq}
%
\begin{equation}\label{Vq}
\hat V_{\bf q} = 2J
\left( \begin{array}{cc}
a & 2 u v  \\
2 u v &  b
\end{array} \right)
\end{equation}

\noindent
with $a = \cos (q_x)$, $b = \cos (q_y)$, 
$u = \cos (q_x / 2)$ and $v = \cos (q_y / 2)$.

At the second stage, the Hamiltonian (\ref{dhamq}) is finally diagonalized to the form
%\makebox{\it \bf dhamdiag}
%
\begin{equation}\label{dhamdiag}
{\cal H} =  \frac{1}{2N} \sum_{n\bf q} \tilde V_{\bf q}^n 
{\sigma}_{\bf q}^n {\cdot \sigma}_{-\bf q}^n , 
\end{equation}
where $ \tilde V_{\bf q}^n = 2J \nu_n({\bf q})$ are the eigenvalues of the matrix 
$V_{\bf q}^{ll'} $ taken with the negative sign,
%\makebox{\it \bf nudef}
%
\begin{equation}\label{nudef}
\nu_1 = 1, \qquad \nu_{2} = - (1 + \cos (q_x) + \cos (q_y)).
\end{equation}
The diagonalizing transformation has the explicit form 
%\makebox{\it \bf Vtrans}
%
\begin{equation}\label{Vtrans}
U_{nl}^{-1}({\bf q}) V^{ll'}_{\bf q} U_{l'n'}({\bf q}) = \tilde V^n_{\bf q} \delta_{nn'} ,
\end{equation}
where the summation over the repeated indices is implied and $\hat U$ is the real unitary matrix,
$\hat U^{-1} = \hat U^T$, i.e., $U_{nl}^{-1} = U_{ln}$.
The columns of the matrix $\hat U$ are the two normalized
eigenvectors $U_n = (U_{1n},U_{2n})$ of the interaction matrix 
$\hat V$:
%\makebox{\it \bf eigenvec}
%
\begin{eqnarray}\label{eigenvec}
&&
U_1  = \sqrt{\frac{1+a}{2+a+b}} \left( - \frac{2uv}{1+a},1 \right),
\nonumber\\
&&
U_2  = \sqrt{\frac{1+b}{2+a+b}} \left( \frac{1+a}{2uv},1 \right).
\end{eqnarray}
The normalized eigenvectors satisfy the requirements of orthogonality 
and completeness, respectively:
%\makebox{\it \bf orthcomp}
%
\begin{equation}\label{orthcomp}
U_{ln}({\bf q}) U_{ln'}({\bf q}) = \delta_{nn'},
\qquad U_{ln}({\bf q}) U_{l'n}({\bf q}) = \delta_{ll'} .
\end{equation}
The Fourier components of the spins ${\bf s}_{\bf q}^l$ and the 
collective spin variables ${\sigma}_{\bf q}^n$ are related by
%\makebox{\it \bf s-sigma}
%
\begin{equation}\label{s-sigma}
{\bf s}_{\bf q}^l = U_{l n}({\bf q}) {\sigma}_{\bf q}^n, 
\qquad {\sigma}_{\bf q}^n = {\bf s}_{\bf q}^l U_{l n}({\bf q}).
\end{equation}
\begin{figure}
\hspace{-0.2cm}
\includegraphics[width=9.5cm]{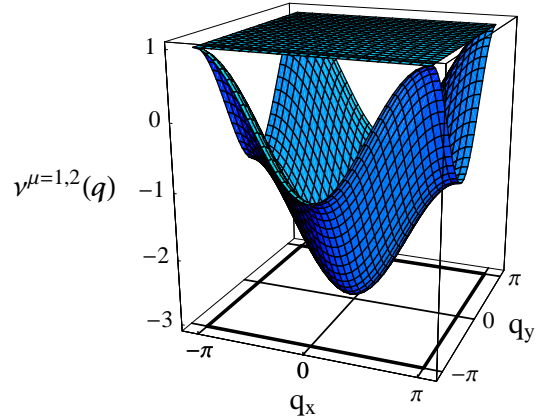}%
\vspace{-0.5cm}
\caption{Reduced eigenvalues of the interaction matrix,
$ \nu_n({\bf q}) = \tilde V_{\bf q}^n/(2J)$ of 
Eq. (\protect\ref{dhamdiag}), plotted over the Brillouin zone.
The flat band signals that half of the degrees of freedom in this
system can rotate freely.}
\label{fig-check-mod}
\end{figure}

The largest dispersionless eigenvalue $\nu_1$ of the interaction matrix 
[see Eq. (\ref{nudef})] manifests frustration in the system which precludes 
an extended short-range order even in the limit $T\to 0$.
Independence of $\nu_1$ of ${\bf q}$ signals that 1/2 of all spin degrees 
of freedom are local and can rotate freely.
The other eigenvalue satisfy
%\makebox{\it \bf nu2qsm}
%
\begin{equation}\label{nu2qsm}
\nu_2 ({\bf q}) \cong -3 + q^2/2
\end{equation}
at small wave vectors, $q^2 \equiv q_x^2 + q_y^2 \ll 1$.
Within our description of the checkerboard lattice, the would be
usual Goldstone mode corresponds to $\nu_2$.
At the corners of the Brillouin zone $(\pm \pi, \pm \pi)$, 
this mode becomes degenerate
with the flat one and defines low energy excitations.
\begin{equation}\label{nu2qborder}
\nu_2 ({\bf \tilde{q}}) \cong 1 - \tilde{q}^2/2
\end{equation}
where $q = (\pm \pi, \pm \pi) + \tilde{q}$, $|\tilde{q}| << 1$.
It is the equivalent of the usual Goldstone mode that is present
in low-dimensional magnets with a continuous symmetry.
Thus, it is this mode that will be associated with a correlation
length, as we will see in section \ref{secCFs}.
It is remarkable that this mode contains all information about
dispersive excitations present in this model.
If we had choosen a four sites unit cell \cite{notreportedhere}
instead of a two sites one, we would have obtained a similar description as the one
done on the pyrochlore antiferromagnet \cite{canalsgaranin00}.
Then, an ``optical'' mode would have been recovered, i.e a mode
with a finite energy gap for all wave vector; the usual Goldstone
mode would have been degenerate with the flat one at $q=0$ and
the lowest not dispersive branch, would have been two times
degenerate.
This clearly shows the similarity of these two lattices as it is 
expected because of the exact equivalence of the local
environment of each site.
The ${\bf q}$ dependences of the eigenvalues $\nu_n$ over the whole Brillouin 
zone are shown in Fig. \ref{fig-check-mod}.\\
In the next section the equations describing spin correlation functions of the
classical checkerboard antiferromagnet in the large-$D$ limit will be obtained with
the help of the classical spin diagram technique.
The readers who are not interested in details can skip to Eq. (\ref{defcfsig})
or directly to Sec. \ref{secThermod}.

\section{Classical spin diagram technique and the large-$D$ limit}
\label{secDinfinity}

The exact equations for spin correlation functions in the limit $D\to\infty$, as well as the $1/D$ corrections,
can be the most conveniently obtained with the help of the classical spin diagram technique.
\cite{garlut84d,gar94jsp,gar96prb} 
A complete description of the technique applied to a non bipartite lattice can be found in 
Ref. \cite{garcan99}.
We only give here an outline of the calculations.

Our goal is to compute spin-spin CF's
\begin{equation}
s^{ll'}({\bf q}) = \langle s_{\bf q}^n s_{-\bf q}^n \rangle
\end{equation}
which are related to CF's of the $\sigma$ variables
\begin{equation}\label{defcfsig}
\sigma^n({\bf q}) = \frac DN 
\langle \sigma_{\alpha \bf q}^n \sigma_{\alpha, {-\bf q}}^n \rangle 
=\frac 1N \langle {\sigma}_{ \bf q}^n {\sigma}_{-\bf q}^n \rangle ,
\end{equation}
through the relation
\begin{equation}\label{defcfs}
s^{ll'}({\bf q}) = U_{l n}({\bf q}) U_{l' n}({\bf q})
\sigma^n({\bf q}) 
\end{equation}
following from Eq. (\ref{s-sigma}).

The analytical expression for the $\sigma$ CF in the SCGA has the Ornstein-Zernike form
%\makebox{\it \bf sigcfOZ}
%
\begin{equation}\label{sigcfOZ}
\sigma^n({\bf q}) = \frac{ D \tilde\Lambda }
{ 1 -  \tilde\Lambda \beta \tilde V_{\bf q}^n } .
\end{equation}
In the large-$D$ limit the expression of $\tilde\Lambda$ simplifies \cite{gar94jsp} and is
given by
%\makebox{\it \bf Lamtil}
%
\begin{equation}\label{Lamtil}
\tilde\Lambda = \frac 2D \frac{ 1 }{ 1 + \sqrt{ 1 + 8L/D } } .
\end{equation}
Here the dispersion $L$ is given by the formula
%\makebox{\it \bf L}
%
\begin{equation}\label{L}
L = \frac{ \tilde \Lambda }{ 2 \cdot 2! } \sum_n
 v_0 \, \!\!\!\int\!\!\! \frac{d{\bf q}}{(2\pi)^d}
 \frac{ ( \beta \tilde V_{\bf q}^n )^2 }
{  1 -  \tilde\Lambda \beta \tilde V_{\bf q}^n }
\end{equation}
The summation $(1/N)\sum_{\bf q}\ldots$ is replaced by the integration
over the Brillouin zone, $v_0$ is the unit cell volume, and $d$ is the spatial
dimensionality.
For the square lattice we have $v_0 = 1$.
The expression for $L$ can be simplified to
%\makebox{\it \bf LbarP}
%
\begin{equation}\label{LbarP}
L = \frac{ \bar P - 1}{ 2\tilde\Lambda }, 
\qquad \bar P \equiv \frac 12 \sum_n P_n ,
\end{equation}
where $P_n$ is the lattice Green function associated with the eigenvalue $n$:
%\makebox{\it \bf Pndef}
%
\begin{equation}\label{Pndef}
P_n =  v_0 \, \!\!\!\int\!\!\! \frac{d{\bf q}}{(2\pi)^d}
 \frac{ 1 }
{  1 -  \tilde\Lambda \beta \tilde V_{\bf q}^n } .
\end{equation}
Now one can eliminate $L$ from Eqs. (\ref{Lamtil}) and (\ref{LbarP}), which
yields the basic equation of the large-$D$ model,
%\makebox{\it \bf sphereq0}
%
\begin{equation}\label{sphereq0}
D \tilde\Lambda \bar P = 1 .
\end{equation}
This nonlinear equation determining $\tilde\Lambda$ as
a function of temperature differs from those considered earlier 
\cite{garlut84d,gar94jsp,gar96jsp,gar96prb} by a more complicated form of
$\bar P$ reflecting the lattice structure.
The form of this equation is similar to that appearing in the theory of the
usual spherical model. \cite{berkac52,joy72pt}
The meanings of both equations are, however, different.
Whereas in the standard spherical model a similar equation account for the pretty
unphysical global spin constraint, Eq.  (\ref{sphereq0}) here is, in fact, the
normalization condition $\langle {\bf s}_{\bf r}^2 \rangle = 1$ for the spin
vectors on each of the lattice sites ${\bf r}$.
Indeed, calculating the spin autocorrelation function in the form symmetrized
over sublattices with the help of Eqs.  (\ref{defcfs}),  (\ref{orthcomp}), and
(\ref{sigcfOZ}), one obtains
%\makebox{\it \bf constr}
%
\begin{eqnarray}\label{constr}
&&
\langle {\bf s}_{\bf r}^2 \rangle =  v_0 \, \!\!\!\int\!\!\! \frac{d{\bf q}}{(2\pi)^d}
\frac 12 \sum_l s^{ll}({\bf q})
\nonumber\\
&&
\qquad
{} =  v_0 \, \!\!\!\int\!\!\! \frac{d{\bf q}}{(2\pi)^d}
\frac 12 \sum_n \sigma^n ({\bf q}) = 
D \tilde\Lambda \bar P .
\end{eqnarray}
That is, the spin-normalization condition is automatically satisfied in our
theory by virtue of Eq.  (\ref{sphereq0}). 
After  $\tilde\Lambda$ has been found from this equation, the spin CFs are readily
given by Eqs. (\ref{sigcfOZ}) and (\ref{defcfs}).

To avoid possible confusions, we should mention that in the paper of Reimers,
Ref. \cite{rei92ga}, where Eq. (\ref{sigcfOZ}) with the bare cumulant  
$\Lambda = 1/D$ has been obtained, the theoretical approach has
been called ``Gaussian approximation (GA)''.
This term taken from the conventional theory of phase transitions based on the
Landau free-energy functional implies that the Gaussian fluctuations of the {\em
order parameter} are considered.
In the microscopic language, this merely means calculating correlation
functions of fluctuating spins after applying the MFA.
Such an approach is known to be inconsistent, since correlations are taken
into account after they had been neglected.
As a result, for the checkerboard lattice one obtains a phase transition at the temperature 
$T_c = T_c^{\rm MFA} = 2J/D$ but immediately finds that the approach breaks
down below $T_c$ because of the infinitely strong fluctuations.
In contrast to this MFA-based approach, the self-consistent Gaussian
approximation used here allows, additionally, to the Gaussian fluctuations of
the {\em molecular field}, which renormalize $\tilde \Lambda$ and lead
to the absence of a phase transition for this class of systems.
The SCGA is, in a sense, a ``double-Gaussian'' approximation; one concerning
fluctuations of the order parameter and the other one describing fluctuations
of the molecular field.

To close this section, let us work out the expressions for the energy and the
susceptibility of the checkerboard antiferromagnet.
For the energy of the whole system, using Eqs. (\ref{dhamdiag}) and (\ref{defcfsig}), as
well as the equivalence of all spin components, one obtains
%\makebox{\it \bf Utot}
%
\begin{equation}\label{Utot}
U_{\rm tot} = \langle {\cal H } \rangle  = -\frac{ N }{ 2 } \sum_n
 v_0\!\!\!\int\!\!\!\frac{d{\bf q}}{(2\pi)^d}
 \tilde V_{\bf q}^n  \sigma^n ({\bf q}).
\end{equation}
To obtain the energy pro spin $U$, one should divide this expression by $2N$.
With the use of Eq.  (\ref{sigcfOZ}), the latter can be expressed through the
lattice Green's function $\bar P$ of Eq. (\ref{LbarP});  then with the help of
Eq. (\ref{sphereq0}) it can be put into the final form
%\makebox{\it \bf U}
%
\begin{equation}\label{U}
U = \frac T2 \left( D - \frac{ 1 }{  \tilde\Lambda } \right).
\end{equation}
The susceptibility pro spin symmetrized over sublattices can be expressed through the
spin CFs as
%\makebox{\it \bf chidef}
%
\begin{equation}\label{chidef}
\chi_{\bf q} = \frac{ 1 }{ 2DT } \sum_{ll'} s^{ll'}({\bf q}) .
\end{equation}
With the use of Eq. (\ref{defcfs}) this can be rewritten in the form
%\makebox{\it \bf chisig}
%
\begin{equation}\label{chisig}
\chi_{\bf q} = \frac{ 1 }{ 2DT } \sum_n W_n^2({\bf q}) \sigma^n({\bf q}),
\qquad W_n({\bf q}) \equiv \sum_l U_{l n}({\bf q}) ,
\end{equation}
where the diagonalized CFs are given by  Eq. (\ref{sigcfOZ}).
From Eq. (\ref{eigenvec}) it follows that in the limit ${\bf q} \to 0$ one has
$W_1=0$ and $W_2=\sqrt{2}$.
Thus the homogeneous susceptibility $\chi \equiv \chi_0$ simplifies to
%\makebox{\it \bf chihomo}
%
\begin{equation}\label{chihomo}
\chi = \frac{ 1 }{ DT } \sigma^2 (0).
\end{equation}
As we shall see in the next section, disappearance of the terms with $n=1$
from this formula ensures the nondivergence of the homogeneous susceptibility
of the checkerboard antiferromagnet in the limit $T\to 0$.
The situation for ${\bf q} \ne 0$ is much more intricate and it will be
considered below in relation to the neutron scattering cross section.

\section{Thermodynamics of the checkerboard antiferromagnet}
\label{secThermod}

To put the results obtained above into the form explicitly well behaved in the
large-$D$ limit and allowing a direct comparison with the results obtained by
other methods for systems with finite values of $D$, it is convenient to use
the mean-field transition temperature  $T_c^{\rm MFA} = 2J/D$ as the energy
scale.
With this choice, one can introduce the reduced temperature $\theta$ and the
so-called gap parameter $G$ according to
%\makebox{\it \bf defthetaG}
%
\begin{equation}\label{defthetaG}
\theta \equiv \frac{ T }{ T_c^{\rm MFA} }, 
\qquad G \equiv \frac D\theta  \tilde\Lambda .
\end{equation}
In these terms, Eq. (\ref{sphereq0}) rewrites as
%\makebox{\it \bf sphereq}
%
\begin{equation}\label{sphereq}
\theta G \bar P(G) = 1 
\end{equation}
and determines $G$ as function of $\theta$.
Here $ \bar P(G)$ is defined by Eq. (\ref{LbarP}), where
%\makebox{\it \bf Pn}
%
\begin{equation}\label{Pn}
P_1 = \frac{ 1 }{ 1 - G } , \qquad 
P_2 =  v_0\!\!\!\int\!\!\!\frac{d{\bf q}}{(2\pi)^d}
 \frac{ 1 }
{  1 -  G \nu_2 ({\bf q}) } ,
\end{equation}
The $\sigma$ CFs of Eq. (\ref{sigcfOZ}), which are proportional to the integrands of
$P_n$, can be rewritten in the form
%\makebox{\it \bf sigcf}
%
\begin{equation}\label{sigcf}
\sigma^n ({\bf q}) = \frac{ \theta G }{  1 -  G \nu_n ({\bf q}) } .
\end{equation}
Further, it is convenient to consider the reduced energy pro spin defined by
%\makebox{\it \bf defUtil}
%
\begin{equation}\label{defUtil}
\tilde U \equiv U/|U_0|, \qquad U_0 = - J,
\end{equation}
where $U_0$ is the energy pro spin at zero temperature.
With the help of Eq. (\ref{U}) $\tilde U$ can be written as
%\makebox{\it \bf Util}
%
\begin{equation}\label{Util}
\tilde U = \theta - 1/G .
\end{equation}
The homogeneous susceptibility $\chi$ of Eq. (\ref{chihomo}) can be rewritten
with the help of Eq. (\ref{nu2qsm})  in the reduced form
%\makebox{\it \bf chitilhom}
%
\begin{equation}\label{chitilhom}
\tilde \chi \equiv 2J \chi = \frac{ G } { 1 + 3G } .
\end{equation}

\begin{figure}
\vspace{0.2cm}
\includegraphics[width=8cm]{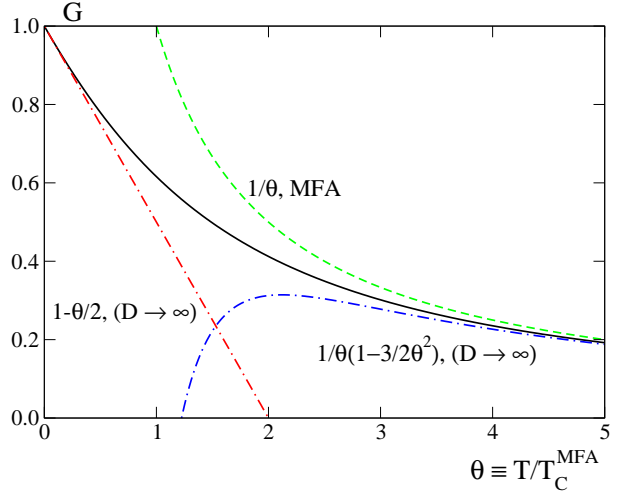}%
\caption{Temperature dependence of the gap parameter $G$ for the checkerboard antiferromagnet.
\vspace{0.cm}
}
\label{check_GvsT}
\end{figure}

The sense of calling $G$ the ``gap parameter'' is clear from Eq. (\ref{sigcf}).
If $G=1$, then the gap in correlation functions closes:
$\sigma^1$ turns to infinity, and $\sigma^2$
diverges at $q\to ( \pm \pi , \pm \pi)$.
For nonordering models, it happens only in the limit $\theta\to 0$,  however. 
The solution of Eq. (\ref{sphereq}) satisfies $G\leq 1$
and goes to zero at high  temperatures. 
If $\theta \ll 1$, the function $\bar P$ is dominated by $P_1 = 1/(1-G)$.
The ensuing asymptotic form of the gap parameter at low temperatures reads
%\makebox{\it \bf GTlo}
%
\begin{equation}\label{GTlo}
G \cong 1 - \frac \theta 2 , \qquad \theta \ll 1 .
\end{equation}
At high temperatures, Eq. (\ref{sphereq}) requires small values of $G$.
Here, the limiting form of $\bar P$ can be shown to be $\bar P \cong 1 + (3/2) G^2$.
The corresponding asymptote of $G$ has the form
%\makebox{\it \bf GThi}
%
\begin{equation}\label{GThi}
G \cong \frac 1 \theta  \left( 1 - \frac{ 3 }{ 2 \theta^2 } \right) , \qquad \theta \gg 1 .
\end{equation}
The numerically calculated temperature dependence of $G$ is shown in
Fig. \ref{check_GvsT}.
Note that in the MFA one has $G=1/\theta$ which attains the value 1 at $\theta=1$.

\begin{figure}
\vspace{0.2cm}
\includegraphics[width=8cm]{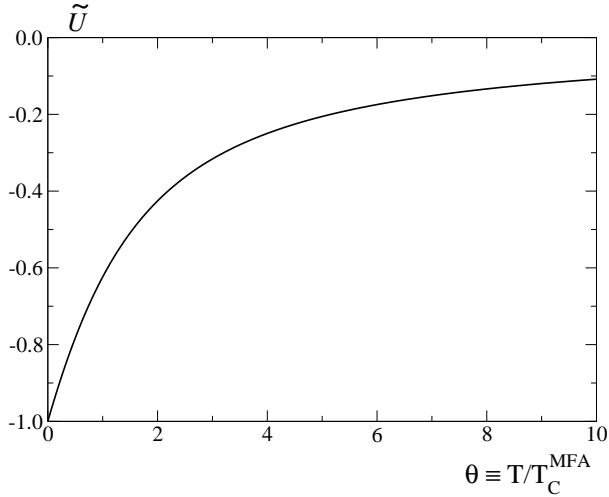}%
\caption{Temperature dependence of the reduced energy for the checkerboard antiferromagnet.}
\label{check_UvsT}
\end{figure}
\begin{figure}
\vspace{0.2cm}
\includegraphics[width=8cm]{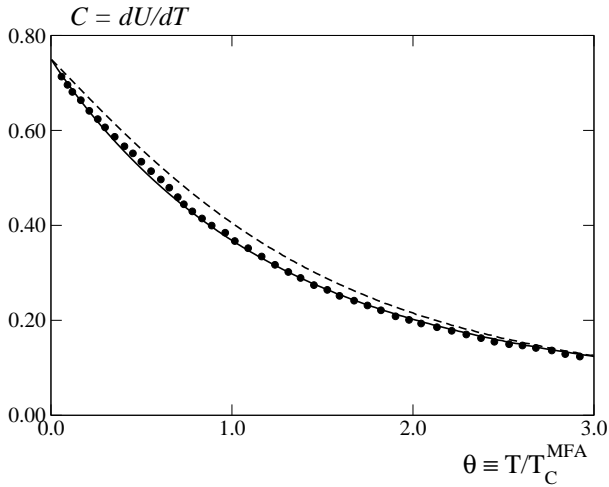}%
\caption{Temperature dependence of the  heat capacity of the checkerboard
antiferromagnet.
The MC results of Ref. \protect\cite{rei92} for the Heisenberg model
($D=3$) on the pyrochlore lattice are represented by circles.
Black line corresponds to the exact solution of the infinite component heisenberg
antiferromagnet on the pyrochlore lattice \cite{canalsgaranin00}.
Dashed line corresponds to the exact solution of the infinite component heisenberg
antiferromagnet on the checkerboard lattice.
}
\label{check_CvsT}
\end{figure}
\begin{figure}
\vspace{0.5cm}
\includegraphics[width=8cm]{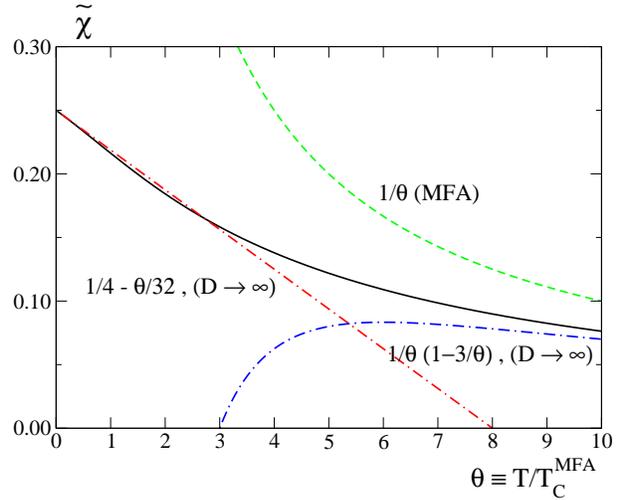}%
\caption{Temperature dependence of the reduced uniform susceptibility of the checkerboard
antiferromagnet.
Asymptotes for high and low temperatures are given, for both the infinite component model
and the mean field case.
}
\label{check_XvsT}
\end{figure}

The temperature dependence of the reduced energy of Eq. (\ref{Util}) is shown
in Fig. \ref{check_UvsT}.
Its asymptotic forms following from Eqs. (\ref{GTlo}) and (\ref{GThi}) are
given by
%\makebox{\it \bf UtilAs}
%
\begin{equation}\label{UtilAs}
\tilde U \cong \left\{
\begin{array}{ll}
 -3/(2\theta), 	& \theta \gg 1 \\
-1 + \theta/2,	& \theta \ll 1 .
\end{array}
\right.	 
\end{equation}
This implies the reduced heat capacity $\tilde C = d\tilde U/d\theta$
equal to 1/2 at low temperatures.
To compare with the MC simulation data of Ref. \cite{rei92} for the heat capacity of the Heisenberg
model we will use, instead of  $\tilde C$, the true heat capacity 
$C = dU/dT = (D/2)\tilde C$ [see Eqs. (\ref{defthetaG}) and (\ref{defUtil})],  which in our approach tends to
$D/4\Rightarrow 3/4$ 
at low temperatures.
The temperature dependance of the heat capacity, compared with previous
MC results of the Heisenberg antiferromagnet on the pyrochlore lattice \cite{rei92}
as well as exact solution of the infinite component classical 
antiferromagnet on the pyrochlore lattice \cite{canalsgaranin00} are shown in 
Fig. \ref{check_CvsT}.
We see that the behavior of these two different models \cite{rei92,canalsgaranin00}
on the pyrochlore lattice are very similar to our results on the checkerboard
lattice, thus confirming the analogy of these two structures despite their
different space dimensionality.

Using Eq. \ref{UtilAs} we compute the low and high temperature asymptotic
behavior of the reduced susceptibility~$\tilde{\chi}$
%\makebox{\it \bf XlowhighT}
%
\begin{equation}\label{XlowhighT}
\tilde \chi \cong \left\{
\begin{array}{ll}
1/4 - \theta /32, 	& \theta \ll 1 \\
(1/\theta)(1 - 3/\theta),	& \theta \gg 1 .
\end{array}
\right.	 
\end{equation}
Its dependence at all temperatures is reported in Fig. \ref{check_XvsT}.
Here also, we recover a behavior very similar to the one of the 
pyrochlore lattice where MC simulations and analytical results have
been obtained for the Heisenberg antiferromagnet \cite{moessnerberlinsky99}
as well as exact results for the infinite component classical 
antiferromagnet \cite{canalsgaranin00}.
The similarity between our results and the local approximation 
\cite{explanation} developped
in Ref. \cite{moessnerberlinsky99} suggest that the checkerboard
antiferromagnet has its thermodynamics governed by local correlations,
as it is expected in a classical spin liquid.

\section{Real-space correlation functions}
\label{secCFs}

The low-temperature behavior of the $\sigma$ correlation
functions of Eq. (\ref{sigcf})  is dominated by their asymptotic
form at small wave vectors, where the would be Goldstone mode
is defined, i.e at $q = (\pm \pi, \pm \pi) + \tilde{q}$, $|\tilde{q}| << 1$.
According to Eqs. (\ref{nudef}), we find,
(\ref{nu2qborder}), and (\ref{GTlo}), by 
%\makebox{\it \bf sigcfTlo}
%
\begin{equation}\label{sigcfTlo}
\sigma^1 \cong 2, \qquad \sigma^2 \cong \frac{2 \kappa^2}{\kappa^2 + \tilde{q}^2}
\end{equation}
where the quantity $\kappa^2=\theta$ in $\sigma^2$ defines the correlation length
%\makebox{\it \bf xidef}
%
\begin{equation}\label{xidef}
\xi_c = \frac 1\kappa = \frac{1}{\sqrt{\theta}} .
\end{equation}
Appearance of this length parameter implies that the real-space spin CFs
defined, according to Eqs. (\ref{defcfs}) and (\ref{fourier}), by
%\makebox{\it \bf cfR}
%
\begin{equation}\label{cfR}
s_{ij}^{ll'} =  v_0\!\!\!\int\!\!\!\frac{d{\bf q}}{(2\pi)^d}
 e^{ i {\rm q \cdot}( {\bf r}_i^l - {\bf r}_j^{l'}) }
U_{l n}({\bf q}) U_{l' n}({\bf q})
\sigma^n({\bf q})
\end{equation}
decay exponentially at large distances at nonzero temperatures.
In contrast to conventional lattices, divergence of $\xi_c$ at $\theta\to 0$
does not lead here to an extended short-range order, i.e., to strong
correlation at distances $r \lesssim \xi_c$.
The zero-temperature CFs are {\em purely geometrical} quantities which are dominated 
by $\sigma^1$ and have the form
%\makebox{\it \bf cfRT0}
%
\begin{eqnarray}\label{cfRT0}
&&
s_{ij}^{ll'} =  2 v_0\!\!\!\int\!\!\!\frac{d{\bf q}}{(2\pi)^d}
 e^{ i {\rm q \cdot}( {\bf r}_i^l - {\bf r}_j^{l'}) }
U_{l 1}({\bf q}) U_{l' 1}({\bf q})
\end{eqnarray}

It is convenient to enumerate CFs by the numbers $n_u$ and $n_v$ defined by 
Eq. (\ref{transl}).
Thus $s_{n_u,n_v}^{ll'}$ is the correlation function of the $l$ sublattice
spin of the ``central'' unit-cell $(0,0)$ with the $l'$ sublattice spin of the
unit-cell translated by $(n_u,n_v)$.
Let us calculate the CFs with $l=l'=1$ at large distances along the 
horizontal line in Fig. \ref{fig-check-and-square}, at small but non
zero temperature.
We use the asymptotic form of the matrix $U_{l n}({\bf q})$ for 
$q = (\pm \pi, \pm \pi) + \tilde{q}$, $|\tilde{q}| << 1$,
as well as the asymptotic form of CFs in Eq. \ref{sigcfTlo}.
This allows to write this correlation function as
\begin{eqnarray}
s_{ij}^{ll'} & = & v_0\!\!\!\int\!\!\!\frac{d{\bf \tilde{q}}}{(2\pi)^d}
 (-1)^n e^{ i n \tilde{q}_x  }
( U_{1 1}^{2}({\bf \tilde{q}}) \sigma^1 + U_{1 2}^{2}({\bf \tilde{q}}) \sigma^2 ) \nonumber \\
 & = & 2 v_0\!\!\!\int\!\!\!\frac{d{\bf \tilde{q}}}{(2\pi)^d}
 (-1)^n e^{ i n \tilde{q}_x  }
( \frac{\tilde{q}_{y}^{2}}{\tilde{q}^2} + 
   \frac{\tilde{q}_{x}^{2}}{\tilde{q}^2} \frac{\kappa^2}{\kappa^2 + \tilde{q}^2} )
\end{eqnarray}
Simplifying this expression and taking into account that the integral of the
cosine over the whole Brillouin zone is zero, one 
arrives at the asymptotic form
%\makebox{\it \bf cf11hor}
%
%
\begin{equation}\label{cf11hor}
\renewcommand{\arraystretch}{2.5}
s_{n,0}^{11} \cong  \frac{ (-1)^n \kappa }{ \pi n } K_1(\kappa n)
\cong  \left\{
\begin{array}{ll}
\displaystyle
 \frac{ (-1)^n }{ \pi n^2 } , 	& \kappa n \ll 1 \\
\displaystyle
\frac{ (-1)^n \kappa^2 }{ \sqrt{2\pi} }  \frac{ e^{-\kappa n} }{
(\kappa n)^{3/2} },	& \kappa n \gg 1,
\end{array}
\right.	 
\end{equation}
where $K_\nu (x)$ is the Macdonald (modified Bessel) function.
The strong decrease of this correlation function with distance even at $T=0$
($\kappa=0$) is not surprising, since our solution spans the whole highly
degenerate ground-state manifold and this degeneracy is not lifted in the
limit $D\to\infty$.
At small but non zero temperature, spin-spin correlations are exponentially
decaying as if interactions were renormalized to zero and drive this 
system to a paramagnetic fix point for finite temperatures.
Let us calculate more generally the CFs with $l=1,2$ and $l'=1,2$  
at large distances along the
horizontal line in Fig. \ref{fig-check-and-square}, at zero temperature.
Following the same previous procedure, we obtain
%\makebox{\it \bf cfgeneralhor}
%
\begin{equation}\label{cfgeneralhor}
s_{n,0}^{11} = s_{n,0}^{22} = \frac{(-1)^n}{\pi n^2}, 
\qquad s_{n,0}^{12} = \frac{(-1)^n}{\pi (n+1/2)^2}
\end{equation}
where $n$ is in unit of the inter-cell distance.
We note that despite the divergence of the correlation length $\xi_c$ when
lowering the temperature, the model do not order, which clearly defines 
this system as a classical two dimensionnal spin liquid.
\begin{figure}
\includegraphics[width=8.5cm]{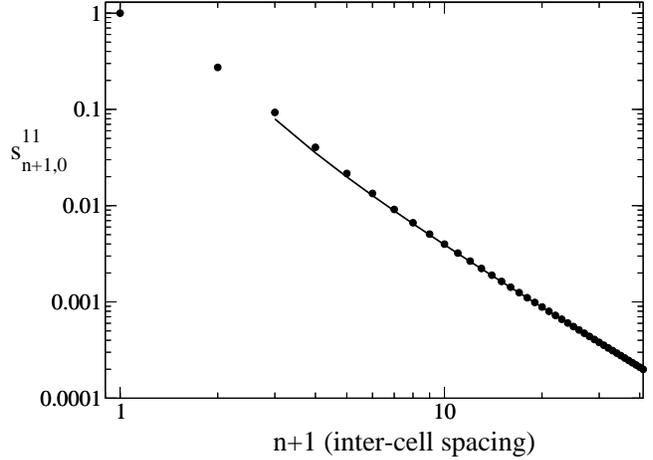}%
\vspace{0.2cm}
\caption{Real-space correlation functions $s_{n,0}^{11}$ at $T=0$
calculated from Eq. (\protect\ref{cfRT0}) along an horizontal
line. 
The distance unit is the interatomic spacing.
The asymptote given by Eq. (\protect\ref{cf11hor}) 
is shown by the dashed lines.
}
\label{check_cf}
\end{figure}

\section{Discussion}
\label{secDiscussion}

In the main part of this article, we have presented in detail the exact
solution for the $D=\infty$ component classical antiferromagnet on the
checkerboard lattice. 
The solution does not show ordering at any 
temperature due to the strong degeneracy of the ground state, and 
thermodynamic functions behave smoothly.
In contrast to conventional two-dimensional magnets, there is no
extended short-range order at low temperature, and $T=0$ is not
a critical point of the system.
Although the correlation associated to the would be Goldstone mode
diverges as $\xi_c \propto T^{-1/2}$, the power law decay 
$\left< s_0 s_r \right> \propto 1/r^2$ at zero temperature of the
spin correlation functions leads to the loss of correlations
at the scale of the interatomic distance.
All these properties characterize this model as a two-dimensional 
classical spin liquid.
As these results are obtained within the $D=\infty$ component 
classical antiferromagnet, it is expected that it mimics the 
large $S$ value Heisenberg antiferromagnet on the checkerboard
lattice.
Surprisingly, there are results \cite{canalswave00,moessner2001} suggesting that
for large $S$, the Heisenberg antiferromagnet  should be ordered on
this lattice.
At this stage, the discrepancy between the different approaches is
not clear.
A possible explanation , as reported in Ref.~\cite{canalswave00},
``is that for large S, there should be quantum 
order by disorder (as supported in \cite{moessner2001}) but this
order by disorder comes with an energy scale of J/S rather than J.
Therefore, in the classical limit, the temperature at which the
correlation length grows exponentially goes to zero.
In other words, the limits $T \rightarrow 0$ and $S \rightarrow \infty$ 
(or $D \rightarrow \infty$) should not commute.''
Nevertheless, it should be stressed that even if there is a discrepancy
at $T = 0$, for finite $T$, we expect the $D=\infty$ case to correctly
describe the checkerboard Heisenberg antiferromagnet as any critical behavior is
expected to appear only in the zero temperature neighbourhood.
A similar problem has been encountered within the present formalism applied
to kagom\'e antiferromagnet \cite{garcan99}.
Here again, finite temperature properties of the infinite component spin
vector model have been found to be close to the $D=3$ Heisenberg case.
However at $T\to0$, the method misses the ``order by disorder'' phenomenon which
selects the coplanar spin manifold.
As a consequence, the $D=3$ zero temperature specific heat is much closer to $1$, 
$C = 11/12 k_{\rm{B}}$, than it would be if really one third of the degrees of
freedom were still fluctuating (as suggested by a mean field analysis).
Wether $1/D$ corrections could resolve this discrepancy is still a work to
be done.

Despite the dimensionality of the checkerboard lattice ($d=2$), there are many
similarities with previously obtained results on the pyrochlore
lattice ($d=3$).
The specific heat has in particular, the same value at zero 
temperature \cite{canalsgaranin00,rei92} {\bf for $D=3$}.
%
%It could be interesting to study the possibility of entropic selection
%in this frame as the description of the lattice is here much more
%simpler than its three-dimensional analog.
%
Susceptibility is also very similar with the one of the Heisenberg
antiferromagnet as well as the infinite component classical
antiferromagnet on the pyrochlore lattice.
MC results, as well as exact and analytical results are very
close to the one obtained here \cite{canalsgaranin00,moessnerberlinsky99}.
This is not surprising as the spin liquid behavior implies that
the lattice is mainly described by local fluctuations.
In fact, it is true at least for finite temperature thermodynamics.
Why the present approach can lead to correct results for the pyrochlore
case and not for the checkerboard case at very low temperatures
has not been addressed in the present work.
We can only note that the common point between the two lattices is the local 
connectivity which is clearly well taken into account within the 
$D \to \infty$ formalism.
Wether dimensionality is relevant is not clear.
At this stage, it can be noted that for the kagom\'e case,
the $D \to \infty$ formalism has missed the reduced specific
heat mechanism (see Ref.~\cite{garcan99,chalker92}) while for the 
pyrochlore case, it has been reproduced (see Ref.~\cite{canalsgaranin00})
Therefore, discrepencies with $D=3$ descriptions could be ascribed to
dimensionality although other subtle phenomena that are not taken into account 
when $D \to \infty$ cannot be ruled out without studying the same model when
including $1/D$ corrections.

To conclude, this classical approach indicates that for infinite component spin
value, the checkerboard antiferromagnet should be disordered and 
behave as a spin liquid.
We give some arguments why the present approach is at variance with results 
of Ref. \cite{moessner2001}.
It is interesting to note that for small spin value, the checkerboard
antiferromagnet is expected to order \cite{canalswave00,moessner2001,fouet2001}.
Therefore the checkerboard antiferromagnet cannot be compared to the pyrochlore
antiferromagnet even if its geometry defines it as its two dimensional analog.
Even if they have the same unit cell, it is probably the global geometry of the
underlying Bravais lattice that drives the physics and not only the local 
connectivity.
%
%

% BibTeX users please use
% \bibliographystyle{}
% \bibliography{}

\begin{thebibliography}{10}
\bibitem{notreportedhere}
The calculations for a four sites unit cell have been done but we
choose to report a two sites description for simplicity. The reader is referred
to previous work if necessary \cite{canalsgaranin00}.
\bibitem{explanation}
Authors have integrated exactly the partition function as well
as the susceptibility of a tetrahedron which also corresponds
in our case to the four sites unit cell of the checkerboard lattice
when described as a square lattice of tetrahedra.
\end{thebibliography}

\end{document}